\documentclass[10pt,a4paper]{article} %removed na a4paper: ,citesort
\usepackage[makeroom]{cancel}
\usepackage{bm}
\usepackage{amsmath}
\usepackage{amsfonts}
\usepackage{graphicx}
\usepackage{nicefrac}
\usepackage{authblk}
\usepackage{cite}
\usepackage[left=2.54cm,top=2.54cm,right=2.54cm,bottom=2.54cm,nohead]{geometry}
\DeclareGraphicsExtensions{.png,.jpg,.pdf}
\usepackage{epstopdf}
\usepackage{float}
\usepackage{subfigure}
\usepackage{fouriernc} 

\title{Theory of fluid slip in charged capillary nanopores}

\makeatletter
\renewcommand\AB@authnote[1]{\textsuperscript{\normalfont#1}}

\makeatother

\author[1]{J. Catalano,}
\author[2]{R.G.H. Lammertink,}
\author[3,4]{P.M. Biesheuvel}
\affil[1]{Department of Engineering, Aarhus University, Hang\o vej 2, 8200 Aarhus, Denmark.}
\affil[2]{Soft matter, Fluidics and Interfaces, MESA$^{+}$ Institute for Nanotechnology, University of Twente, Drienerlolaan 5, 7522 NB Enschede, The Netherlands.}
\affil[3]{Wetsus, European Centre of Excellence for Sustainable Water Technology, Oostergoweg 9, 8911 MA Leeuwarden, The Netherlands.}
\affil[4]{Laboratory of Physical Chemistry and Soft Matter, Wageningen University, Dreijenplein 6, 6703 HB Wageningen, The Netherlands.}

\date{} %remove date

\begin{document}

\def\yPh{\vphantom{j_{\mathrm{charge},x}}}
\newcommand{\jch}{\overline{j_\text{ch} \yPh}}
\newcommand{\ux}{\overline{u \yPh}}

\renewcommand{\t}{\widetilde}
\newcommand{\s}[1]{\mathrm{_{#1}}}

\maketitle

\begin{abstract}
Based on the capillary pore model (space-charge theory) for combined fluid and ion flow through cylindrical nanopores or nanotubes, we derive the continuum equations modified to include wall slip. We focus on the ionic conductance and streaming conductance, cross-coefficients of relevance for electrokinetic energy conversion and electro-osmotic pumping. We combine the theory with a Langmuir-Stern 1-pK charge regulation boundary condition resulting in a non-monotonic dependence of the cross-coefficients on salt concentration.
\end{abstract} 

\section{Introduction}

Charged capillary nanopores, or nanotubes, allow for combined water and ion transport, and have applications in desalination and energy conversion~\cite{Weinstein_science,Post_Environ,Achilli_JMS,Mattia}. Continuum theories for flow of fluid and ions describe the ions as ideal point charges and the water as an unstructured fluid. The mathematical theory combines the Navier-Stokes, Nernst-Planck and Poisson equations~\cite{Dresner}. For pores that are much longer than wide, Osterle and co-workers~\cite{Morrison&Osterle,Gross&Osterle,Fair&Osterle} introduced the capillary pore model, or space charge (SC) theory, which is based on assuming chemical and mechanical equilibrium in the direction transverse to flow, i.e., in the $r$-direction in a cylindrical geometry. Using this simplification, calculation of transport in the flow direction can be discoupled from solution of the Poisson-Boltzmann equation in $r$-direction. This theory has been shown to accurately predict ion fluxes and fluid flow, not only in track-etched pores and highly ordered structures, but also in materials such as ion-selective membranes that are characterized by a much lower degree of order~\cite{Cwirko&Carbonell_JMS}. Based on the SC theory, further simplifications can be made by assuming a homogenized potential and concentration in $r$-direction (``uniform potential''-model), and/or by neglecting concentration gradients in axial direction~\cite{Sonin,HawkinsCwirko&Carbonell,Verbrugge,Bowen&Welfoot,Tedesco,Catalano_JPCM}.

\begin{figure}[H]%[!ht]%[H]
	\centering
	\includegraphics[scale=0.67]{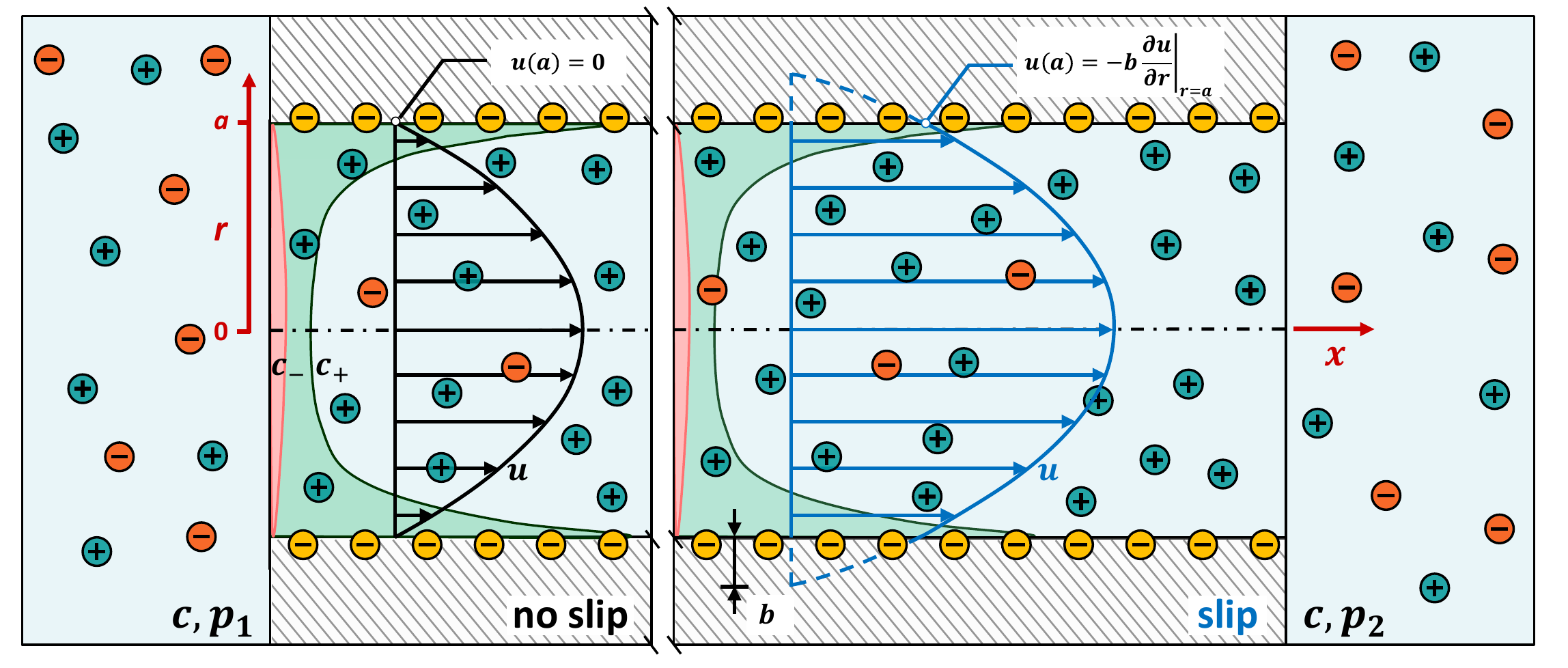}
	\caption{Schematic representation of the velocity profile due to a pressure difference $\Delta p=p_1-p_2$ in a nanopore with and without slip. In this example, the pore is negatively charged and connects two reservoirs at the same salt concentration $c$. The slip length is approximately $b/a=1/4$. Anion and cation concentrations in the pore are denoted with red and blue colors.}\label{fig_schem}
\end{figure}

One outstanding question in the description of flow through nanopores and nanotubes is the relevance of fluid slip at pore walls, expressed mathematically as the slip length (see Fig.~\ref{fig_schem}). This slip length is an inverse measure of the friction between water and pore wall. For non-wetting fluids transported through (carbon) nanotubes (CNTs), very high slip lengths are reported, up to order of 1 $\mu$m  and larger~\cite{Holt_science,Majumder_Nature}, implying that fluid can flow almost without friction through CNTs. 

In the case of slippage on hydrophilic surfaces limited experimental work has been reported, which most probably is the result of the  difficulties associated with its experimental determination. For hydrophilic surfaces the slip length is orders of magnitude lower than for hydrophobic surfaces. For hydrophobic materials, slip lengths have been reported up to hundreds of nanometers, for instance for oxygen plasma-treated PDMS surfaces~\cite{Huang_JFM}. Instead, for charged pores or tubes, and with water as electrolyte, the slip length is assumed to be much smaller~\cite{Joly_JCP}. In any event, regardless of the nature of the surface, the presence of slip on the pore walls can have drastic effects in micro- and nanofluidic systems in a variety of processes: e.g., transport of particles in solution~\cite{Ajdari}, enhanced  heat transfer~\cite{Haase_JFM}, and electrokinetic processes~\cite{Ren&Stein,Joly_JCP,Davidson}. Here we will show that even a slip length of the order of 1 nm can significantly modify key parameters relevant for performance of membranes containing charged nanopores, either for energy harvesting, energy conversion or desalination.

For planar slits, fluid wall slip was described by Ren and Stein~\cite{Ren&Stein}, while pressure-driven electrokinetic flow was analyzed for asymmetric zeta potentials~\cite{Jing&Bhushan} and for fluids with non-Newtonian viscoelastic properties~\cite{Berli&Olivares,Matin&Khan}. Recently, Rankin and Huang analyzed the SC theory for capillary (cylindrical) pores including fluid wall slip~\cite{Rankin_Langmuir}. 
  
In the present work we modify the capillary pore model of Osterle and co-workers to include fluid wall slip. We will consider both the case of fixed wall charge, as well as  wall charge that responds via a surface ionization equilibrium to the local electric potential. We will present results for alumina pores, but the results have general validity for other charged porous membranes as well. 
%We do restrict the derivation to ``hard'' materials (i.e., we exclude polyelectrolyte coated pores). 

\section{The capillary pore model}

In the present manuscript, we solve the capillary pore model for situations that the salt concentration is equal on both sides of the pore, and thus leave out as driving force a concentration difference across the membrane. (Note that this simplifying assumption may not be exactly correct: even with an overall zero concentration difference across the pore, the (virtual) concentration can still change through the pore, with only the values at either pore end fixed at the same value.) %Because we consider equal salt concentrations, the presented theory does not describe membrane desalination or energy harvesting from water salinity differences. 
This simplifies the SC theory significantly because we do not need to consider axial gradients in concentration, chemical potential, or osmotic pressure, all being zero. We can also replace ``virtual'' concentration $c_v$ from ref.~\cite{Peters} by $c$, which is the salt concentration in the external bath. The virtual total pressure $p_{t,v}$ can be replaced by hydrostatic pressure $p$ because there are no gradients in the osmotic pressure. 
%where we left out subscripts $h$ for ``hydrostatic'', and $v$ for ``virtual''. The same we do for the ``axial'' electric potential $\phi$.
%
The derivation of the capillary pore model, or space charge (SC) theory, closely resembles that of Osterle and co-workers \cite{Gross&Osterle,Fair&Osterle}, and is also discussed in refs.~\cite{Sashidar&Ruckenstein,Peters}. 
%We refer to ref.~\cite{Peters} for a detailed treatment of the equations in this section. 

Central to the SC-theory are the Navier-Stokes (NS) equation including local ionic charge, the Nernst-Planck (NP) equation including ion advection, and the Poisson-Boltzmann (PB) equation solved in radial direction. For the development of these equations, see ref.~\cite{Peters}. Different from ref.~\cite{Peters}, we now include the possibility of fluid slip at the pore walls, i.e.,
\begin{equation}\label{no-slip}
\t{u}(a)=-b \> \left. \frac{\partial \t{u}}{ \partial \t{r}}\right|_{\tilde{r}=a}
\end{equation}
where $\t{u}$ is the fluid velocity in $x$-direction [because all velocities and fluxes are in $x$-direction, we will leave out subscript $x$], $a$ is the pore radius, and $b$ is the slip length, inversely related to the fluid-wall friction. 
\newpage
We solve the PB-equation in $r$-direction ($r=\t{r}/a$), which results in
\begin{equation} \label{eq:PB}
\frac{1}{r}\frac{\partial}{\partial r}\left(r\frac{\partial \psi}{\partial r}\right) = \frac{c}{\lambda^2_\text{ref}}\sinh\psi
\end{equation}
where $0<r<1$, and where
\begin{equation}
\lambda\s{ref}=\frac{1}{a} \> \sqrt{\frac{\varepsilon \Phi_B}{2 F \t{c}\s{ref}}}
\end{equation}
is a dimensionless reference Debye length in units of the cylinder radius $a$ and dimensional reference concentration $\t{c}\s{ref}$ $\text{ }(\varepsilon=\varepsilon\s{w} \cdot \varepsilon_0)$. %\emph{called Debye ratio in Probstein 1973}. 
The dimensionless concentration is $c=\t{c}/\t{c}\s{ref}$. The $r$-dependent dimensionless electric potential $\psi$ can be multiplied by the thermal voltage $\Phi \s{B}=R \s{g} T/F$ to obtain a dimensional voltage. The boundary conditions of the PB-equation are
\begin{equation}\label{eq:BoundaryPB}
\left.\frac{\partial\psi}{\partial r}\right|_{r=0} = 0 \hspace{1.2em} , \quad \left.\frac{\partial\psi}{\partial r}\right|_{r=1} = +\sigma
\end{equation}
where the dimensionless wall charge $\sigma$ recalculates to the dimensional wall charge $\t{\sigma}$ (in C/m$^2$) by
\begin{equation}
\sigma=\frac{\t{\sigma}\, a}{\varepsilon \> \Phi \s{B}}.
\end{equation}
The dimensionless parameter $\alpha$ to be used further on is given by 
\begin{equation}
\alpha= \frac{\mu \s{w} D}{\t{c}_\mathrm{ref}R \s{g} T a^2}
\end{equation}
where $\mu \s{w}$ is the fluid (water) viscosity, $D$ the ion diffusion coefficient when both ions have the same diffusion coefficients, and otherwise is given by $D=\sqrt{D \s{+} \, D \s{-}}$ where $+$ and $-$ refer to cation and anion, respectively~\cite{Catalano_JPCM}.

Based on $r$-dependent fluxes (directed in the $x$-direction), the next step is to derive the radially averaged flux, which for a flux $j(r)$ takes the form
\begin{equation}\label{eq:poreaverage}
 \overline{j} = 2\int^1_0 j(r) \, rdr.
\end{equation}

For the dimensionless ionic current in the $x$-direction we obtain (compare with Eq. (24) in ref.~\cite{Peters})
\begin{equation} \label{eq:j_ch}
j_\mathrm{ch}(r)  = -2c \, \sinh\psi \>\> u(r)-2c \, \cosh\psi^* \frac{\partial\phi}{\partial x}
\end{equation}
where we have left out the term dependent on the gradient in chemical potential. The current $j_\mathrm{ch}$ can be multiplied by $F \t{c} \s{ref} \> D / \ell $, where $\ell$ is the length of the pore (or, tube), to arrive at a current density in A/m$^2$. The dimensionless axial potential $\phi$ can be multiplied with $\Phi \s{B}$ to obtain a dimensional voltage. The function $\psi^*$ is given by
\begin{equation}\label{eq:psi_star}
\psi^*=\psi+\ln\sqrt{\frac{D_-}{D_+}}
\end{equation}
which for equal ion diffusion coefficients simplifies to $\psi^* = \psi$. The case of unequal ion diffusion coefficients is also considered in refs.~\cite{Gross&Osterle,Sashidar&Ruckenstein}.

An analytical expression for the dimensionless fluid velocity, $u(r)=\t{u}(r) \, \ell / D$, is obtained on the basis of Eq. (31) of ref.~\cite{Peters}, where we leave out the chemical potential-dependent term, and make use of \mbox{$u(1)=-b/a\>\left.\partial_r u\right|_{r=1}$}, to arrive at
%
%\begin{align}
%\begin{split}
%\alpha \left( u(r) - u(1) \right) =-\frac{1}{4} (1-r^2)\cdot\frac{\partial p_{h}}{\partial x} 
%-2\lambda_\mathrm{ref}^2 (\psi(r) - \psi_\mathrm{wall}) \frac{\partial\phi}{\partial x}
%\end{split}
%\end{align}
%
%\medskip
%
%\begin{align}
%\begin{split}
% \frac{\partial u(r)}{\partial r}=\frac{1}{2\alpha} r \frac{\partial p_{h}}{\partial x} - \frac{2 \lambda_\mathrm{ref}^2}{\alpha} \frac{\partial\psi(r)}{\partial r} \frac{\partial\phi}{\partial x}
%\end{split}
%\end{align}
%
%\begin{align}
%\begin{split}
%\alpha u(r) + \alpha b \left[
%\frac{1}{2\alpha}  \frac{\partial p_{h}}{\partial x} - \frac{2 \lambda_\mathrm{ref}^2}{\alpha} \sigma \frac{\partial\phi}{\partial x}\right]  =-\frac{1}{4} (1-r^2)\cdot\frac{\partial p_{h}}{\partial x} 
%-2\lambda_\mathrm{ref}^2 (\psi(r) - \psi_\mathrm{wall}) \frac{\partial\phi}{\partial x}
%\end{split}
%\end{align}
%
\begin{equation}\label{eq:u_r}
\alpha \cdot u(r)=-\left[ \frac{1}{4} (1-r^2)+\frac{b}{2a}\right]\cdot\frac{\partial p}{\partial x} 
-2\lambda_\mathrm{ref}^2 \left[\psi - \psi_\mathrm{w}- \frac{b}{a} \sigma \right] \cdot \frac{\partial\phi}{\partial x}
\end{equation}
where the dimensionless pressure $p$ can be multiplied by $\t{p} \s{ref}=\t{c} \s{ref} \, R \s{g} \, T$ to arrive at a pressure with unit Pa. For $b=0$, Eq. (\ref{eq:u_r}) is the same as Eq. (27) in ref.~\cite{Morrison&Osterle}. The potential at the wall is $\psi \s{w}$.
Inserting $u(r)$ in Eq. (\ref{eq:poreaverage}), the pore-averaged fluid velocity, $\ux$, becomes
%
%\begin{align}
%\begin{split}
%\alpha \ux =-\int_0^1 \left( \frac{r}{2} (1-r^2)+\frac{b}{a}r\right) dr \cdot\frac{\partial p_{h}}{\partial x} 
%-4\lambda_\mathrm{ref}^2 \int_0^1\left( r(\psi(r) - \psi_\mathrm{wall})- \frac{b}{a} \sigma r \right) dr \cdot \frac{\partial\phi}{\partial x}
%\end{split}
%\end{align}
%
%\begin{align}
%\begin{split}
%\alpha \ux =- \left[ \frac{r^2}{4} - \frac{r^4}{8} +\frac{b r^2}{2a} \right]^1_0 \cdot\frac{\partial p_{h}}{\partial x} 
%-4\lambda_\mathrm{ref}^2 \left[ \int_0^1 (\psi(r) - \psi_\mathrm{wall}) rdr - \frac{b \sigma}{2a} \right] \cdot \frac{\partial\phi}{\partial x}
%\end{split}
%\end{align}
%
\begin{align}
\begin{split}
\alpha \cdot \ux =- \frac{1}{8}\left(1 +4\frac{b}{a} \right) \cdot\frac{\partial p}{\partial x} 
-4\lambda_\mathrm{ref}^2 \left[ \int_0^1 \left( \psi - \psi_\mathrm{w} \right) rdr - \frac{b}{a}\frac{\sigma}{2} \right] \cdot \frac{\partial\phi}{\partial x}
\end{split}
\end{align}
%
%\medskip
%
\begin{figure}[H]%[!ht]%[H]
	\centering
	\includegraphics[scale=0.40]{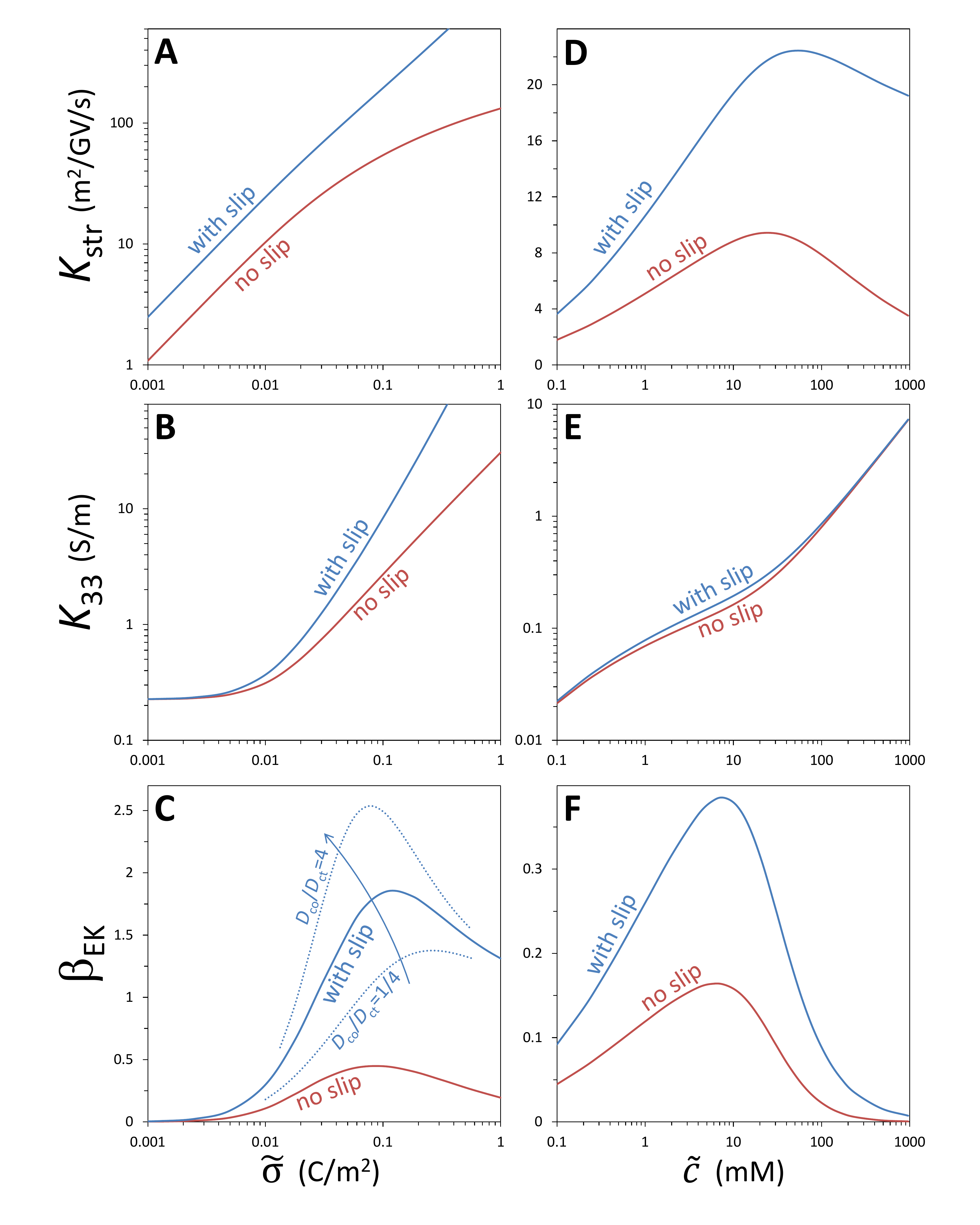}
	\caption{Streaming conductance $K \s{str}$, ionic conductance $K_{33}$, and figure-of-merit $\beta \s{EK}$ for a capillary pore with and without fluid wall slip. Panels A-C show results as function of wall charge (salt concentration $\t{c}=30$ mM), and panels D-F for alumina as function of $\t{c}$. Panel C shows results for unequal coion vs. counterion diffusion coefficient. See main text for parameter settings.}\label{fig}
\end{figure}

\newpage

\noindent while for $\jch$ we arrive at
%
%\begin{equation} \label{eq:j_ions}
%j_\mathrm{ch}(r)  = - 2 c \sinh\psi \> u(r) - 2 c \cosh\psi \frac{\partial\phi}{\partial x}.
%\end{equation}
%
%\begin{equation}
%\begin{split}
%j_\mathrm{ch}(r)  = - 2 \frac{c}{\alpha} \sinh\psi \left[-\left[ \frac{1}{4} (1-r^2)+\frac{b}{2a}\right]\cdot\frac{\partial p_{h}}{\partial x} 
%-2\lambda_\mathrm{ref}^2 \left[(\psi - \psi_\mathrm{wall})- b/a \sigma \right] \frac{\partial\phi}{\partial x} \right] \\ 
%- 2 c \cosh\psi \frac{\partial\phi}{\partial x}.
%\end{split}
%\end{equation}
%
\begin{equation}
\jch  = 4 \frac{c}{\alpha} \int_0^1 \left( \sinh\psi \left( \frac{1}{4} -\frac{r^2}{4}+\frac{b}{2}\right) \cdot \frac{\partial p}{\partial x}
 + 2 \left(\lambda_\mathrm{ref}^2 \, \sinh\psi  \left( \psi - \psi_\mathrm{w} - \frac{b}{a} \sigma \right) -\frac{\alpha}{2}\cosh\psi^*\right) \cdot \frac{\partial\phi}{\partial x} \right) rdr
\end{equation}
%
%\begin{equation}
%\jch  = \int_0^1 4 \frac{c}{\alpha} \sinh\psi \left[ \frac{1}{4} (1-r^2)+\frac{b}{2}\right] rdr \cdot\frac{\partial p_{h}}{\partial x}
% + 8 \frac{c}{\alpha} \int_0^1 \left( \sinh\psi \lambda_\mathrm{ref}^2 \left( \psi - \psi_\mathrm{wall} - b/a \sigma \right) -\frac{\alpha}{2}\cosh\psi\right) rdr \cdot \frac{\partial\phi}{\partial x} 
%\end{equation}
%\begin{equation}
%\begin{split}
%\jch  = \int_0^1 4 \frac{c}{\alpha} \sinh\psi \left[ \frac{1}{4} (1-r^2)+\frac{b}{2}\right] rdr \cdot\frac{\partial p_{h}}{\partial x} \\
% + 8 \frac{c}{\alpha} \int_0^1 \left( \sinh\psi \lambda_\mathrm{ref}^2 \left( \psi - \psi_\mathrm{wall} - b/a \sigma \right) -\frac{\alpha}{2}\cosh\psi\right) rdr \cdot \frac{\partial\phi}{\partial x} 
%\end{split}
%\end{equation}
%
and thus the four (dimensionless) cross-coefficients $L_{11}$, $L_{31}$ and $L_{33}$ become %~\cite{Rankin_Langmuir}
\begin{equation}\label{eq:Lmatrix}
\begin{alignedat}{2}
L_{11}&=+\,\hspace{0.35em}\frac{1}{8\alpha}&\cdot&\left(1+4 \frac{b}{a}  \right)\\
L_{13}&=+\frac{4\lambda_\mathrm{ref}^2}{\alpha} &\cdot & \int^1_0  \left(\psi - \psi_\mathrm{w}-\frac{b}{a}\sigma \right) rdr = L \s{str} = \\
=L_{31}&=- \hspace{0.7em}\frac{c}{\alpha}&\cdot&\int_0^1  \sinh\psi \left( 1-r^2+2 \frac{b}{a} \right) rdr \\
L_{33}&=-\,\hspace{0.4em}\frac{8 c}{\alpha} &\cdot&  \int^1_0  \left(\lambda_\mathrm{ref}^2 \, \sinh\psi \left( \psi - \psi_\mathrm{w} -\frac{b}{a}\sigma \right)-\frac{\alpha}{2}\cosh\psi^*\right) rdr
\end{alignedat}
\end{equation}
where $L_{11}$ is the intrinsic water permeability and relates fluid velocity to pressure drop. The two off-diagonal coefficients, $L_{13}$ and $L_{31}$, are equal due to Onsager symmetry and relate fluid flow $u$ to potential drop $\Delta\phi$, and at the same time relate current $I$ to pressure drop $\Delta p$. Following refs.~\cite{Heyden_PRL,Ren&Stein}, we will call these off-diagonal coefficients the ``streaming conductance'', $L \s{str}$ (the same parameter is called ``electro-osmotic permeability'' in refs. \cite{Chakravarti_JMS, Haldrup_AcaNano}). Finally, the ionic conductance $L_{33}$ relates current to a voltage difference. Eqs. (\ref{eq:Lmatrix}) must be equivalent to Eqs. (21)-(26) in ref.~\cite{Rankin_Langmuir} though the appearance is somewhat different.

The coefficients $L \s{ij}$ can be recalculated to the dimensional parameters $K \s{ij}$ by
\begin{equation}\label{eq:Kvalues}
\begin{alignedat}{3}
K_{11}&=L_{11}&&\cdot \frac{D}{\t{c} \s{ref} \, R \s{g} T} =\frac{a^2}{8\mu \s{w}}\cdot \left(1+4\frac{b}{a}\right)\quad&&\left[\frac{\text{m$^2$}}{\text{Pa$\cdot$s}}\right]\\
K \s{str}&=L \s{str}&&\cdot \hspace{1.1em}\frac{D}{\Phi \s{B} } \quad\quad&&\hspace{0.23em}\left[\frac{\text{m$^2$}}{\text{V$\cdot$s}}\right] \\
K_{33}&=L_{33}&&\cdot \hspace{0.3em}\frac{F\,D\,\t{c} \s{ref}}{ \Phi \s{B}} \quad\quad&&\hspace{0.54em}\left[\frac{\text{S}}{\text{m}}\right].
\end{alignedat}
\end{equation}

Note that these expressions for the $K$-coefficients are based on the open area of straight pores. To describe a true membrane, the right-hand sides in Eq. (\ref{eq:Kvalues}) must be multiplied by membrane porosity $\epsilon$ and divided by tortuosity $\tau$.

\section{Surface ionization}

In a typical calculation of charged pores or capillaries, wall charge $\t{\sigma}$ is set to a fixed value. However, many materials have ionizable surface charge which responds to local pH. This local pH depends on the pH in the external bath, pH$_\infty$, and on the potential at the location of the ionizable wall groups (potential relative to that in the external bath), $\psi_0$ \cite{Heyden_PRL,Biesheuvel_JPCM, Secchi}. For an amphoteric material such as alumina or titania (which can be positively or negatively charged, dependent on pH), a ``Langmuir-Stern'' 1-pK model is applicable which leads to~\cite{Biesheuvel_JCIS,Trefalt_Langmuir}

\begin{equation}\label{eq:ioniz}
\t{\sigma}=e\cdot N\cdot\left(\frac{1}{2}-\frac{1}{1+10^{\text{pK}-\text{pH}_\infty}\cdot\exp(-\psi \s{0})}\right)
\end{equation}
where $N$ is  the number of ionizable sites, for alumina about 5 per nm$^2$, i.e., $e\cdot N \sim800$ mC/m$^2$. Furthermore, for alumina, pK $\sim 9-9.5$ (titania $\sim 5-5.5$). Eq. (\ref{eq:ioniz}) depends on $\psi_0$ which is the potential at the material's very surface, where the ionizable material reacts with protons/hydroxyl ions~\cite{Koopal_EA}. This potential is different from, and in magnitude larger than, the potential at the ``start of the diffuse layer'', i.e., the potential at the Stern plane, which is often denoted by $\psi \s{d}$. This Stern potential is the potential at the outer boundary of the diffuse layer. In the diffuse layer the potential profile is described by the PB-equation. For the Stern potential we used $\psi \s{w}$ in the previous section. % ($\psi \s{wall}$ in ref.~\cite{Peters}). 
The two potentials, $\psi \s{w}$ and $\psi \s{0}$, are related by the Stern (layer) capacity, which is the capacity of the uncharged ``dielectric'' layer in between locations ``0'' and ``w'', according to
\begin{equation}\label{eq:sigma_stern}
\t{\sigma}=C \s{St}\>\Phi \s{B}\left( \psi \s{0}-\psi \s{w} \right)
\end{equation}
where $C \s{St}$ is the Stern capacity, typically of the order of $C \s{St}\sim1$ F/m$^2$. Note that in the framework of this theory of the Gouy-Chapman-Stern electric double layer, both the Stern \emph{layer} and Stern \emph{plane} are free of adsorbed ions, and are uncharged~\cite{Koopal_EA}. The Stern layer separates the charged surface (at location ``0'') from the Stern plane, which is the starting point for the (charged) diffuse layer that extends into solution. 

Eqs. (\ref{eq:ioniz}) and (\ref{eq:sigma_stern}) can be combined to 
\begin{equation} \label{eq:mixedBC}
\left(\frac{1}{2}-\frac{\sigma \, \varepsilon \, \Phi \s{B}}{e\,N\,a}\right)^{-1}-1-10^{\text{pK}-\text{pH}_\infty}\cdot\exp\left(-\frac{\sigma\,\varepsilon}{C \s{St}\, a}-\psi \s{w}\right)=0
\end{equation}
which is an implicit relation between $\psi \s{w}$ and $\sigma$ to be used as mixed boundary condition in the numerical solution for the PB-equation.
Eq. (\ref{eq:mixedBC}) can be simplified to
\begin{equation}
\left(\frac{1}{2}-f_1 \cdot\sigma\right)^{-1}-1-f_2\cdot\exp\left(-f_3 \cdot\sigma-\psi \s{w}\right)=0
\end{equation}
where $f_1=\varepsilon\Phi \s{B}/eNa$, $f_2=10^{\text{pK}-\text{pH}_\infty}$, and $f_3=\varepsilon/C \s{St}a$.

\section{Results and Discussion}

%\emph{reverse in the figure K13 with K33, ie K33 will be A and D ???}

In this section we present calculation results based on the following parameter settings. In all cases, the pore has a radius of $a=5$ nm (pore diameter $d \s{p}=10$ nm). Unless otherwise noted, the salt concentration is $\t{c}=30$ mM. For calculations with equal diffusion coefficients, for both ions, $D=1\cdot 10^{-9}$ m$^2$/s. The water viscosity is $\mu \s{w}=8.9\cdot 10^{-4}$ Pa$\cdot$s. For the reference concentration $\t{c} \s{ref}$ we use the convenient value of 1 mM, but the actual choice is arbitrary. In case of wall slip, the slip length is set to $b=1.25$ nm, thus $b/a=1/4$. We only use this single value and do not analyze the slip length-dependence in more detail. We also neglect a possible dependence of $b$ on wall charge as suggested in ref.~\cite{Joly_JCP} and used in ref.~\cite{Jing&Bhushan} for describing the pressure-driven flow in charged slits with wall slip. Parameters for the ionization model are given below. Finally, we use a temperature of $T=298$ K, and the dielectric constant of water is set to $\varepsilon \s{w}=78$. 

%\subsection{fixed charge capillary nanotubes}

For nanopores or nanotubes with fixed wall charge, one may expect that both streaming conductance, $K \s{str}$, and ionic conductance, $K_{33}$, increase with surface charge, and indeed Fig.~\ref{fig}(A,B) shows this expectation to be correct. Furthermore, with slip, the $K$-coefficients are always significantly higher (for a certain charge), except for $K_{33}$ at low charge where the presence of slip makes no difference. The clear enhancement of ionic conductance might be profitable in membrane-based electrochemical processes, for instance redox flow batteries which work at high molarity ($\geq 1$ M), where a decrease in membrane ionic resistance results in higher battery current density.

Though these coefficients are important to consider, for electrokinetic energy conversion (EKEC), we are particularly interested in the ``figure of merit'', $\beta \s{EK}$, which uniquely determines the energy conversion efficiency, and is one of the key factors to determine the generated electric power~\cite{Catalano_JPCM}. This factor depends on $K_{11}$, $K \s{str}$, and $K_{33}$, and was predicted to increase indefinitely with wall charge density according to an analysis based on the ``uniform potential model'', see Eq. (21) in ref.~\cite{Catalano_JPCM}. However, as we show in Fig.~\ref{fig}(C), in contrast to that prediction, $\beta \s{EK}$ not only levels off at high charge, but beyond an optimum charge starts to decrease strongly. In this particular calculation, the optimal wall charge is around 60 mC/m$^2$ for the no-slip case, and no more than 80 mC/m$^2$ in the case with wall slip. Beyond a wall charge of $\sim 200$ mC/m$^2$, $\beta \s{EK}$ decreases again. The window of the predicted high $\beta \s{EK}$ is in agreement with calculations for planar slits in ref.~\cite{Ren&Stein} where the maximum efficiency was predicted at $5-30$ mC/m$^2$ for $h=100$ nm slits,  $b/h=0.1-0.5$ and $\t{c}=1$ $\mu$M. Similarly, the experimental results for the nitrocellulose/SPS membrane system~\cite{Haldrup_AcaNano} in a 30~mM LiCl solution showed a maximum EKEC efficiency at surface charge densities of $\sim 100-250$ mC/m$^2$ for $\sim 8-10$ nm pore diameters. 

%\emph{the same optimum vs wall charge also given by Ren and Stein, 2006, fig 5, for a planar slit, with the maximum in the range 5-30 mC/m2}

%\subsection{capillary nanotubes with ionizable surface charge} \label{section:ioniz}

Shifting our attention to materials with an ionizable surface charge, we consider as an example alumina, described by the Langmuir-Stern (LS) model of Eq. (\ref{eq:ioniz}) with parameter settings $N=5$ nm$^{-2}$ and $C \s{St}=0.1$ F/m$^2$, while we use a value of $\text{pK}-\text{pH}_\infty=\pm2$ (e.g., pH$_\infty$ 7 and pK$ \s{alumina}$ 9). As shown by Secchi \emph{et al.}~\cite{Secchi}, for ionizable materials, the ionic conductance, $K_{33}$, will not level off at low salt concentration, as in Fig.~\ref{fig}(B), but continues to decrease. For the LS-surface ionization model incorporated in SC theory, we find a similar result, though with a quite non-monotonic dependence of $K_{33}$ on salt concentration, see Fig.~\ref{fig}(E). %, which is plotted on log-log scale. 
In this case, between salt concentrations of 0.1 and 100 mM we cannot derive a unique power dependence of $K_{33}$ on salt concentration, as was possible in the theory of ref.~\cite{Secchi}. The dependence of the slip length on the ionic conductance is weak or absent.

Whereas $K_{33}$ monotonically increased with salt concentration for a material like alumina, as we will show next, this is not the case for the streaming conductance, $K \s{str}$, see Fig.~\ref{fig}(D), and neither for the figure-of-merit, $\beta \s{EK}$, see Fig.~\ref{fig}(F). Interestingly, the salt concentration that maximizes $\beta \s{EK}$ is around \mbox{$6-8$ mM}, a salt concentration where the Debye length is about equal to the pore radius of the pores in this particular calculation.

Finally, we have analyzed the effect of coion diffusion coefficient over counterion diffusion coefficient (the coion is the ion with the same charge sign as the pore wall), $D \s{co} / D \s{ct}$. As shown in Fig.~\ref{fig}(C) and in agreement with a prediction in ref.~\cite{Catalano_JPCM}, we find that an increase in this ratio (while keeping \mbox{$D \s{avg}=\sqrt{D \s{co} D \s{ct}}$}$ \\ 
=1\cdot 10^{-9}$ m$^2$/s the same) increases the figure-of-merit, $\beta \s{EK}$.

\section*{Conclusions}

The capillary pore model of Osterle and co-workers was modified to account for pore wall slip, and the cross-coefficients were derived in the special case, relevant for EKEC applications, of a zero concentration difference between the two ends of the capillary. Following the capillary pore model we quantified the EKEC performance in terms of the figure-of-merit, $\beta \s{EK}$. We showed that the presence of even moderate slip on the pore walls of the nanoscopic capillary, with a slip length of the order of 1 nm, results in significant enhancements of the streaming and ionic conductances. This in turn positively affects $\beta \s{EK}$. 
%As an example we reported calculations of $\beta \s{EK}$ for capillaries with different wall charge with and without slip on the pore walls. 
For the maximum in $\beta \s{EK}$, which is achieved at a relatively low wall charge for both the no-slip and with-slip cases, we found a 4-fold increase in $\beta \s{EK}$ when slip is included. We also showed that a further increase in $\beta \s{EK}$ can be achieved by selecting coions and counterions with high and low diffusion coefficient, respectively. 
The capillary pore model was combined with the Langmuir-Stern isotherm to represent ionizable surfaces. It could be concluded that similar to the case of a fixed charge, slip increases $\beta \s{EK}$ significantly. Additionally, the presence of ionizable surfaces resulted in a non-monotonic dependence of streaming conductance and ionic conductance on salt concentration. 

\section*{Acknowledgments} 
Part of this work was performed in the cooperation framework of Wetsus, European Centre of Excellence for Sustainable Water Technology (www.wetsus.eu). Wetsus is co-funded by the Dutch Ministry of Economic Affairs and Ministry of Infrastructure and Environment, the Province of Frysl\^an, and the Northern Netherlands Provinces. Support was also provided by the Villum Foundation (Grant No. VKR022356, Young Investigator Programme) and The Aarhus University Research Foundation.

\end{document}